\documentclass{osa-article}

\journal{opticajournal}

\usepackage{braket}

\begin{document}

	\title{Compact Polarization-Entangled Photon Source Based on Coexisting Noncritically Birefringent and Quasi Phase Matching in a Nonlinear Crystal}
	
	\author{Chun-Yao Yang\authormark{1,$\dagger$}, Chao-Yuan Wang\authormark{1,$\dagger$,$\ddag$}, Kuan-Heng Lin\authormark{1}, Tsung-Ying Tsai\authormark{1}, Chih-Chia Lin\authormark{1}, Carlota Canalias\authormark{2}, Li-Bang Wang\authormark{1}, Atsushi Yabushita\authormark{3}, and Chih-Sung Chuu\authormark{1,*}}
	
	\address{\authormark{1}Department of Physics and Center for Quantum Science and Technology, National Tsing Hua University, Hsinchu 30013, Taiwan\\
 
             \authormark{2}Department of Applied Physics, Royal Institute of Technology, Roslagstullsbacken 21, 114 21 Stockholm, Sweden\\

	         \authormark{3}Department of Electrophysics, National Yang Ming Chiao Tung University, Hsinchu 30010, Taiwan }
	
	\email{\authormark{$\dagger$}These authors contribute equally to this work.\\
              \authormark{$\ddag$}cywang91403@gmail.com\\
	       \authormark{*}cschuu@phys.nthu.edu.tw} 
	
	
	
	\begin{abstract*}
Polarization-entangled photons are indispensable to numerous quantum technologies and fundamental studies. In this paper, we propose and demonstrate a novel source that generates collinear polarization-entangled photons by simultaneously achieving two distinct types of phase-matching conditions (noncritically birefringent and quasi phase matching) in a periodically poled nonlinear crystal with 
a large poling period of 2 mm. The photon pairs are generated in a polarization-entangled state with a fidelity and concurrence of 0.998 and 0.935, respectively, and violate the Clauser-Horne-Shimony-Holt inequality by 84 standard deviations. The compact source does not require interferometer, delicate domain structures, or post selection, and is advantageous for scalable quantum computing and communication, where many replicas or chip-scale devices are needed.
\end{abstract*}

	\section{Introduction}
        Spontaneous parametric down conversion (SPDC) is a popular, powerful, and easily controlled method for generating strongly correlated (entangled) photon pairs. The stable performance of SPDC 
        in a room temperature environment allows the generated photon pairs to be utilized as a practical source for fundamental studies and applications of quantum information \cite{Aspect82, Briegel98, Knill01}. Among various degrees of freedom possibly entangled in the SPDC photon pairs, polarization entanglement has been widely and successfully used to demonstrate quantum information applications, such as the quantum key distribution (QKD) \cite{Ekert91,Bennet92}, quantum teleportation \cite{Bouwmeester97,Sherson06, Sun16,Chen08}, dense coding \cite{Mattle96,Williams17}, entanglement swapping \cite{Pan01,Goebel08,Ma12}, and entanglement distillation \cite{Kwiat01,Wu19}. Many remarkable works have been done for years as above mentioned. However, scalable quantum technologies \cite{Kimble08} still requires the development of compact entangled-photon sources, of which the quantum properties are insensitive to possible fabrication defects.
	  \\\indent

To generate polarization-entangled photons, previous works usually require multiple nonlinear crystals or poling structures, interferometers, and post selection, thus increasing the complexity or suffering from the poling errors and efficiency loss. For example, critical phase matching in $\beta$-phase barium borate (BBO) \cite{Kwiat95} has been widely used to generate biphotons in two non-overlapping cones. However, due to the noncollinear geometry and short pump coherence length, the collection of a fraction of photon pairs emitted at the intersections as well as short crystals are necessary to generate
polarization-entangled photons, which reduces the generation rate. Periodic poling, a well-developed technique nowadays for achieving various quasi-phase matching (QPM) conditions \cite{Fejer92}, is also commonly used with nonlinear crystals to generate biphotons at desired wavelengths. Unlike the BBO crystals, multiple crystals \cite{Pelton04,Steinlechner12} or multiple poling structures \cite{Ueno12}
are usually needed for these crystals to create the polarization-entangled states. To avoid the multiple crystals or poling structures, interferometers \cite{Kwiat94,Fiorentino04} based on the Sagnac setup \cite{Kim06,Fedrizzi07} or beam displacers \cite{Fiorentino08} may be employed but at the cost of requiring interferometric alignment. 
Interference at a beam splitter \cite{Ou88,Shih88} has also been used for creating polarization-entangled degenerate photons by post selection (discarding half of the generated photon pairs). Nevertheless, the degeneracy of the photon pairs can be disadvantageous for connecting hybrid quantum systems. \\

In this paper, we propose and demonstrate a novel polarization-entangled photon source utilizing two distinct types of phase matching, the noncritically birefringent phase matching (NBPM) and QPM with a large poling period of 2 mm, in a PPKTP crystal. Under the NBPM condition, the KTP crystal generates collinear photon pairs, which simplifies the beam alignment and experimental setup as compared to the critical phase matching condition. The well-developed periodic poling technique for the KTP crystals allows us to control the QPM conditions and improve the efficiency of the SPDC process at the desired wavelengths. The large poling period also greatly reduces the effect of fabrication error on the generated photon pairs. By achieving both the NBPM and QPM conditions in the same crystal, we demonstrate a compact source of polarization-entangled photons. \\
     
     This paper consists of two parts as follows. In the first part, we present the design of the PPKTP crystal that can directly generate polarization-entangled photons without the need of interferometers, multiple poling structures, or post selection. In the second part, we first characterize the generation of biphotons under the NBPM conditions. The quantum properties of the generated photon pairs are evaluated by measuring the anticorrelation parameters and the Franson-type interference. We also measure the joint spectral power density of the photon pairs and compare it with the theoretical prediction, which shows the feasibility for generating non-degenerate photon pairs. We then experimentally demonstrate the generation of polarizarion-entangled photons and evaluate their entanglement by the violation of the Clauser-Horne-Shimony-Holt (CHSH) inequality and the quantum state tomography (QST).


	\section{Theory}

	    The effective Hamiltonian of a type-II nonlinear optical parametric process in the interaction picture is \cite{Rubin94,Keller97}
	    \begin{equation}
            H_{1}=\epsilon_{0}\int_{V}dV\chi_{psi}E^{(+)}_{p}E^{(-)}_{s}E^{(-)}_{i}+ {\rm H.c.},
            \end{equation}
	    where \textit{V} is the volume of the crystal, $\chi_{psi}$ is the component of the nonlinear susceptibility tensor, with the indices \textit{p}, \textit{s}, \textit{i} representing the polarization directions of the pump, signal, and idler fields, respectively, and H.c. means the Hermitian conjugate. The signal (idler) fields could be written as
            \begin{equation}
            \qquad E^{(-)}_{r}=\int d\omega_{r}\frac{e_{\omega_{r}}}{n_{r}(\omega_{r})}a^{\dagger}_{r}(\omega_{r})e^{-i[k_{r}(\omega_{r})z-\omega_{r}t]},\quad r= s, i.
            \end{equation}
     Here $e_{\omega_{r}}=\sqrt{\hbar\omega_{r}/2\epsilon_{0}V_{Q}}$, 
     $V_{Q}$ is the quantization volume, and $a^{\dagger}_{r}(\omega_{r})$ 
     is the photon creation operator. By considering a strong, monochromatic ($\omega_{p}=\omega_{s}+\omega_{i}$), plane-wave pump field  
     and defining the phase mismatch as $\Delta k=k_{p}-k_{s}-k_{i}$,
    the entangled-photon state can be obtained as follows,
            \begin{equation}
	    \ket{\psi}=\int d\omega_{s}d\omega_{i}dz\, A_{si}(\omega_{s},\omega_{i})\chi_{psi}e^{i\Delta kz} a^{\dagger}_{s}(\omega_{s})a^{\dagger}_{i}(\omega_{i}) \ket{0}. 
	    \end{equation}
            We see that the phase mismatch $\Delta k$ affects the efficiency and bandwidth of SPDC, which limits the available frequencies of the generated photon pairs. Nevertheless, it can be improved by QPM in periodically poled crystals. Such degree of freedom also plays an important role in our proposed scheme. \\

	    We consider a periodically poled nonlinear crystal with a duty cycle \textit{D} and poling period $\Lambda$. The effective spatially varying nonlinear susceptibility can be described by the Fourier series,
	    \begin{equation}
	    \begin{aligned}
	    \chi^{\rm eff}_{psi}(z)=\chi_{psi,0}+\sum_{m\neq0}\chi_{psi,m} \ e^{i\pi mD}e^{-i\frac{2m\pi }{\Lambda}z}
	    \end{aligned}
	    \end{equation}
        where $\chi_{psi,0}=\chi_{psi}(2D-1)$ and $\chi_{psi,m}=\chi_{psi}2D \mathrm{sinc}(\pi mD)$.
		The effective susceptibility varies with \textit{D}, which can be controlled in the fabrication process, and enables us to achieve NBPM and QPM simultaneously. Assume that the pump light is polarized along the \textit{y}-axis of a PPKTP crystal. The generated biphoton state is then given by
	    \begin{equation}
            \ket{\psi}=\int d\omega_{s}d\omega_{i}\alpha_{yzy}(\omega_{s},\omega_{i})\left(
     \ket{z}_{\omega_s}\ket{y}_{\omega_i}+\frac{\beta_{yyz}(\omega_{s},\omega_{i})}{\alpha_{yzy}(\omega_{s},\omega_{i})})\ket{y}_{\omega_s}\ket{z}_{\omega_i}\right)
            \end{equation}
		By choosing proper amplitude ratio for the two participating phase matching conditions, we can thus generate the SPDC state in a polarization-entangled state. \\

To obtain the maximally entangled state in our scheme with a single poling structure and a single crystal, it is important to control the efficiency of each phase matching condition. These efficiencies are proportional to the square of the $m$th order nonlinear susceptibility $\chi^{\rm eff}_{psi,m}$, 
which depends on the duty cycle of the poled crystal as follows,
	    \begin{equation}
	    \begin{aligned}
            &R_{0}\;\propto\;{\chi^{2}_{psi,0}}=\chi_{psi}^{2}(2D-1)^{2},\\
	    &R_{m}\;\propto\;{\chi^{2}_{psi,m}}=\chi_{psi}^{2}[2D{\rm sinc}(\pi mD)]^{2},
            \end{aligned}
	    \end{equation}
	    where $R_{0}$ and $R_{m}$ correspond to the NBPM and the $m$th order QPM. 
     Therefore, by designing the duty cycle \textit{D}, the spectra and efficiencies of the NBPM and QPM can be equalized to generate the maximally entangled state. In our previous work \cite{Yang16}, a poled crystal with irregular duty cycle has been used for shaping the spectrum of the generated biphotons. Note that, compared to the QPM with a duty cycle of 0.5, the generation efficiency will be reduced by a factor of $1/[\pi D {\rm sinc}(\pi D)]^2$ in order to achieve the same efficiency for the NBPM and QPM. \\

		The feasibility of the proposed scheme is confirmed by numerical calculation for the case when NBPM is accompanied with the first order ($m=1$) QPM. The NBPM process generates photon pair with an \textit{z}-polarized signal photon and a \textit{y}-polarized idler photon with angular frequencies of $\omega_{s,z}$ and $\omega_{i,y}$, respectively. The QPM process generates a 
        \textit{y}-polarized signal photon and an \textit{z}-polarized idler photon with angular frequencies $\omega_{s,y}$ and $\omega_{i,z}$, respectively. In practice, we let \textit{y} = \textit{H} and \textit{z} = \textit{V}, where \textit{H} and \textit{V} are the horizontal and vertical polarizations, respectively. In Fig. 1(a) we calculate the poling periods $\Lambda$ of the QPM that generate the biphotons (satisfying $\omega_{s,V}=\omega_{s,H}$ and $\omega_{i,V}=\omega_{i,H}$) with the wavelengths given in Fig. 1(b) as a function of the pump wavelengths.
	    \begin{figure}[t]
	    	\graphicspath{{fig/}}
	    	\includegraphics[width=0.7\textwidth]{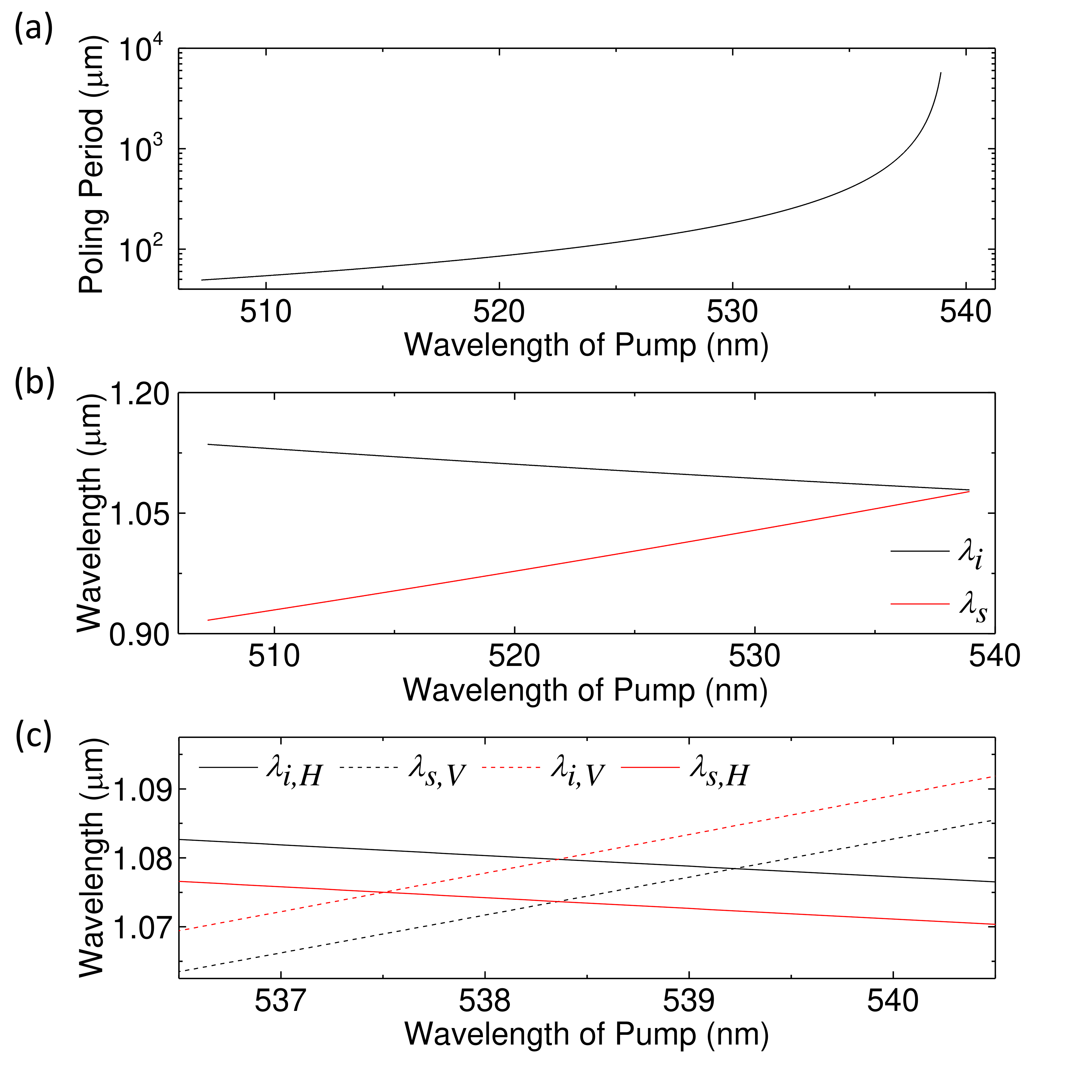}
	    	\setlength{\belowcaptionskip}{-0cm}
	    	\footnotesize
	    	\centering
	    	\caption{
				(a) The poling periods of a PPKTP crystal that can generate entangled photons are shown as a function of the pump wavelength.
				(b) Signal wavelength $\lambda_{s}$ and idler wavelength $\lambda_{i}$ of the entangled photons in (a) as the function of the pump wavelength.
				(c) Signal wavelength $\lambda_{s,V} (\lambda_{s,H})$ and idler wavelength $\lambda_{i,H} (\lambda_{i,V})$ generated by the NBPM (black lines) and QPM (red lines) processes when the poling period of the PPKTP crystal is 2 mm.
	    	}
	    	\label{figure:induction loop}
		\end{figure}
			Figure 1(c) shows the wavelengths of the signal photons $\lambda_{s,V} (\lambda_{s,H})$ and idler photons $\lambda_{i,H} (\lambda_{i,V})$ generated by NBPM (QPM) process using a PPKTP crystal with a poling period of 2 mm, which indicates that polarization-entangled photons can be generated at the pump wavelength of 538.4 nm with the signal and idler wavelengths at 1073.7 nm and 1079.8 nm, respectively. The frequency nondegeneracy of the entangled photons allows them to be separated by a frequency-dependent splitter. Moreover, the signal and idler photons are generated in collinear geometry. \\


        Finally, we briefly discuss the tolerance of the designed crystal to the fabrication error of the periodic poling. The normalized conversion efficiency can be expressed as \cite{Fejer92}
        \begin{equation}
        \eta=\frac{1}{N^{2}} \vert\sum^{N}_{j=1}\exp(-i\Phi_{j})\vert^{2},
        \end{equation} 
        where \textit{N} is the total number of domains and 
        \begin{equation}
        \Phi_{j} =\Delta k\delta z_{j}+\delta\Delta kz_{j,0}
        \end{equation}
        is the accumulated phase error on the $j$th 
        boundary between the domains due to the phase mismatch $\Delta k$ at the position $z_{j}$. It can be seen that when the phase is matched or the domain structure is perfectly fabricated for the $m$th order 
        QPM, $\Phi_{j}$ would be zero or $\pi$ so that the conversion efficiency $\eta$ is 1.\\ 
        \begin{figure}[t]
        	\graphicspath{{fig/}}
        	\includegraphics[width=1\textwidth]{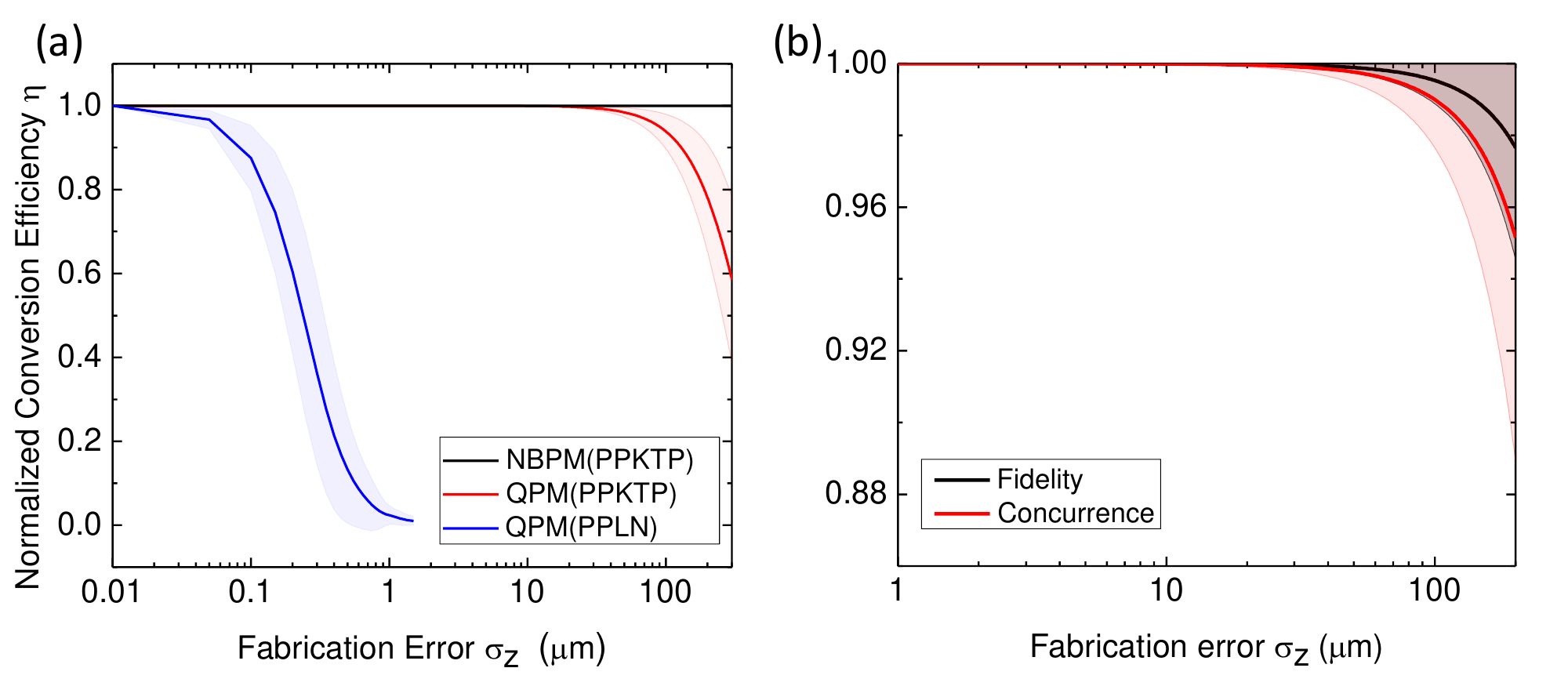}
        	\setlength{\belowcaptionskip}{-0cm}
        	\footnotesize
        	\centering
        	\caption{(a) Normalized conversion efficiencies with various fabrication errors. The black and red curves correspond to the NBPM and the first-order ($m=1$) QPM, respectively, in a PPKTP crystal. The blue curve corresponds to the first-order ($m=1$) QPM in a PPLN crystal with a poling period of 15 $\mu$m. The light red shaded area represents the standard deviation at a given error $\sigma_{z}$ in the Monte Carlo simulation. (b) Concurrence and fidelity of the entangled states for different fabrication errors. The thick lines represent the averages over 2000 samples. The shaded areas show one standard deviation around the averages.}
        	
        	\label{figure:induction loop}
        \end{figure}
      \\\indent
        To quantify the effect of the fabrication error, we firstly assume that $\delta\Delta k= 0$ so that the phase error mainly comes from the boundaries at the incorrect positions and consider the duty cycle defect, which is one of the most common error occurring during the fabrication process. Since the crystal in our design only has 8 domains, we applied the Monte Carlo simulation to calculate the normalized conversion efficiency. We randomly choose different position errors at each boundary and keep the average of the errors to be zero. The average normalized conversion efficiencies for the NBPM and first-order QPM are shown in Fig. 2(a). The curve (red) of QPM shows that the designed large poling period effectively alleviate the influence of the fabrication defect as compared to a crystal with a poling period of $\sim 15\ \mu$m (blue curve). The efficiency stays beyond 90$\%$ even if the boundary position error goes up to 100~$\mu$m. This advantage also helps the generation of polarization-entangled photons since the coefficients $\alpha_{yzy}$ and $\beta_{yyz}$ in Eq. (6) are determined by the efficiency. The concurrence and fidelity of the generated entangled state are also simulated with the same method, of which the result is shown in Fig. 2(b). Both concurrence and fidelity stay high even if the fabrication errors are larger than 100 $\mu$m.
          
	\section{Experimental results}
 	\subsection{Biphoton generation with noncritically birefringent phase matching (NBPM)}
        The experimental setup for generating photon pairs using a KTP crystal under the NBPM condition is shown in Fig. 3. 
         \begin{figure}[t]
        	\graphicspath{{fig/}}
        	\includegraphics[width=1\textwidth]{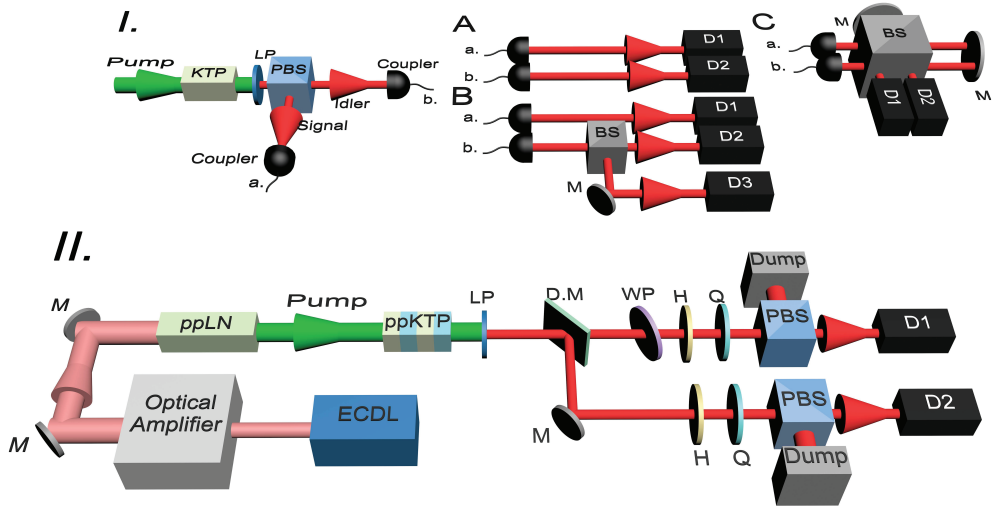}
        	\setlength{\belowcaptionskip}{-0cm}
        	\footnotesize
        	\centering
			\caption{(I) Experimental setup for generating entangled photon pair with the NBPM condition. The generated photon pairs were guided into fiber couplers and sent to three different setups for evaluating their quantum properties: The measurements of (A) $\alpha_{2d}$, (B) $\alpha_{3d}$, and (C) the Franson-type interference. (II) Experimental setup for generating and verifying the polarization-entangled photons. The wavelength-tunable pump light is generated by the second harmonic generation (SHG) of an external cavity diode laser (ECDL). The polarization-entangled photons were generated by a single PPKTP crystal with a single poling structure. Each photon pair are spatially separated by a dichroic mirror (DM) and analyzed by two sets of half-wave plate (HWP, H), quarter-wave plate (QWP, Q), polarizing beam splitters (PBSs), and single-photon counting modules (SPCMs). M stands for mirrors.
        	}
        	\label{figure:induction loop}
        \end{figure} 
		To exclude the QPM, we prepare a 8-mm-long unpoled Rb-doped KTP (RKTP) crystal to generate SPDC photon pairs. A 
        horizontally polarized pump light from a 532-nm cw laser (Coherent Inc., Verdi G5-SLM) was confocally focused into the RKTP crystal satisfying the type-II phase matching condition to generate photon pairs consisting of vertically polarized signal photons and horizontally polarized idler photons. The generated photon pairs were coupled into multimode fibers and detected by single-photon counting modules (SPCMs). The coincidence count of the photon pair was measured using a time-to-digital converter (FAST ComTec, MSC6) as a function of delay time between the signal and idler photons. The pair rate was measured as 1,350 ${\rm s^{-1} mW^{-1}}$ per bandwidth ($\sim$3 nm) at a pump power of 120 $\mu$W, implying a brightness of $R_sR_i/R_c \simeq 3.56\times10^6$ ${\rm s^{-1} mW^{-1}}$, where $R_{c}$, $R_{s}=1.2\times10^4$ s$^{-1}$, $ R_{i}=1.72\times10^4$ s$^{-1}$ are the count rates of the coincidences (pair counts), signal photons, and idler photons, respectively. \\


	     The coincidence measurements were exploited to estimate the quantum properties of the generated photon pairs. We firstly examine the nonclassical correlation of the generated photon pairs by measuring the anticorrelation parameter $\alpha_{2d}=R_{c}/(\tau_{c}R_{s}R_{i})$ \cite{Grangier86}, where $\tau_{c}$ is the coincident window. A polarizing beam splitter (PBS) is used to separate the signal and idler photons so as to be detected by two SPCMs [see Fig. 3(I.A)]. 
	    \begin{figure}[t]
	    	\graphicspath{{fig/}}
	    	\includegraphics[width=0.8\textwidth]{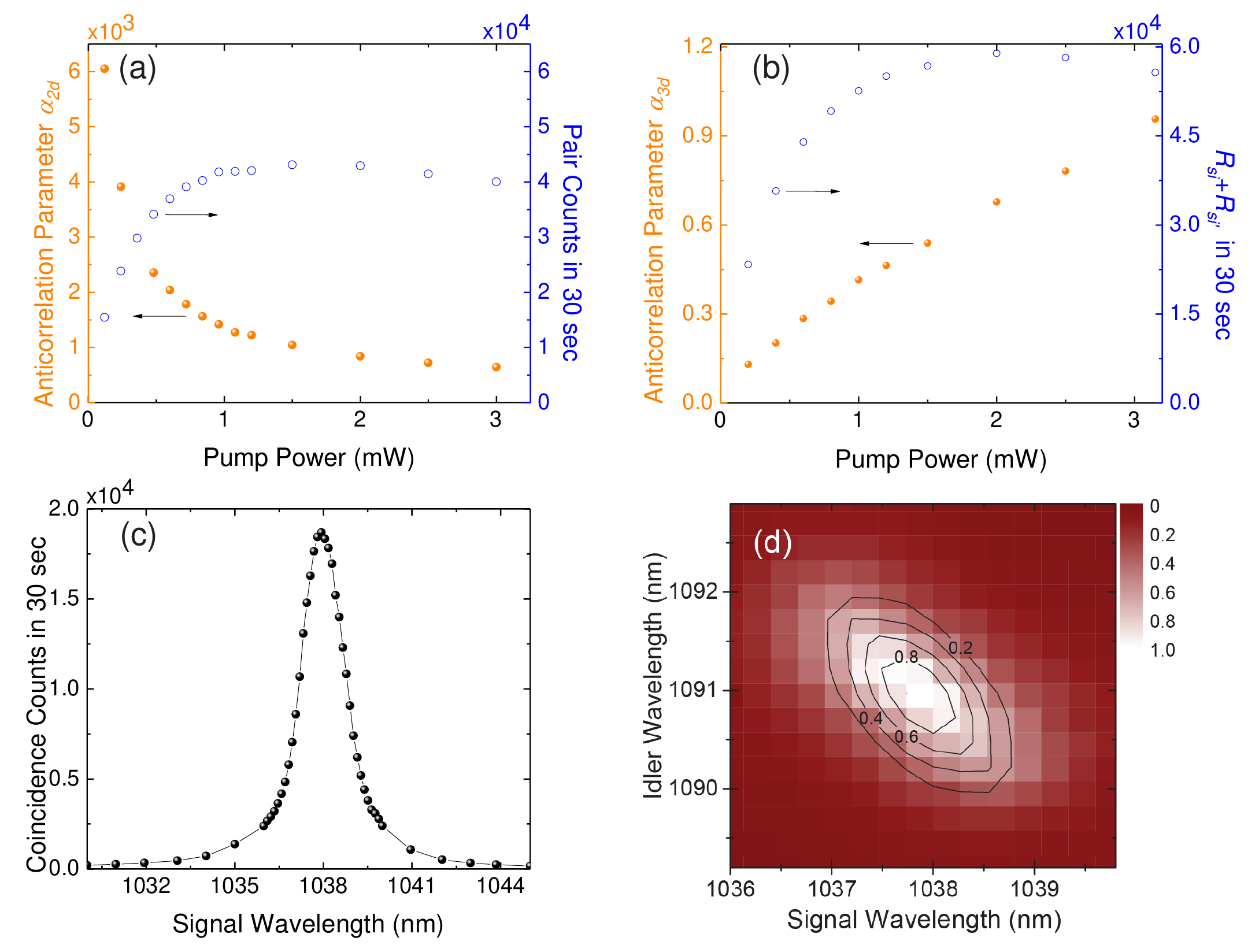}
	    	\setlength{\belowcaptionskip}{-0cm}
	    	\footnotesize
	    	\centering
	    	\caption{(a) Measured anticorrelation parameter $\alpha_{2d}$ (filled circles) and pair counts (open circles) as a function of pump power. (b) Measured anticorrelation parameter $\alpha_{3d}$ (filled circles) and $R_{si}+R_{si'}$ (open circles) as a function of pump power. (c) Measured spectral power density of the signal photons. (d) Measured joint spectral power density of the signal and idler photons. The contour is the calculated joint spectral power density which takes into account the finite bandwidth (about 1 nm) of the tunable bandpass filter. The saturation of count rates in (a) and (b) are due to the long dead time (10 $\mu$s) of the single-photon detectors used in these measurements, in which the single counts are about 50,000 s$^{-1}$.
	    	}
	    	\label{figure:induction loop}
		\end{figure}
	    Figure 4(a) shows the measured $\alpha_{2d}$ as a function of the pump power. The measured values of $\alpha_{2d}$ exceed the classical limit of unity, confirming the nonclassical correlation between the signal and idler photons in each pair. \\

	    We also verify the antibunching character of the heralded single photons by measuring the anticorrelation parameter $\alpha_{3d}=R_{sii^{'}}R_{s}/(R_{si}R_{si^{'}})$ with a Hanbury-Brown-Twiss (HBT) interferometer, where $R_{sii^{'}}$ represents the coincidence rate between three detectors. For the measurement of $\alpha_{3d}$, the system shown in Fig. 3(I.A) was modified by inserting a beam splitter (BS) in the optical path of the idler photons followed by two SPCMs [see Fig. 3(I.B))], in which the detection of a signal photon is treated as a trigger. At low pump power as shown in Fig. 4(b), the fact that $\alpha_{3d}<0.5$ shows the single photon nature or antibunching character of the heralded idler photons.\\

		The time-energy entanglement of the signal and idler photons was evaluated by a Franson-type interferometer \cite{Franson89} shown in Fig. 3(I.C). The Franson-type interferometer is composed of two spatially-separated Michelson interferometers, each with two imbalanced arms. By adjusting the arm length differences in the signal and idler paths, the interference between the biphoton wavefunctions passing through the long-long and short-short arms can be observed. The fringe visibility is measured to be $98\pm0.5$\%, manifesting the time-energy entanglement by violating the classical limit of 50\%.\\

		To confirm the feasibility of generating non-degenerate photon pairs, we measure the spectral power density of the signal photons and the joint spectral power density of the photon pairs, which are shown in Figs. 4(c) and 4(d), respectively. The spectral power densities are recorded by placing a tunable bandpass filter (TBF) with a 1-nm bandwidth in front of the SPCM in the beam paths of the signal or idler photons. The finite bandwidth of the filter results in a hardly observable side peaks in the joint spectral power density, which is proportional to square of a sinc function centered at where the NBPM is perfectly satisfied. Nevertheless, the measured joint spectral power density shows that the generation of nondegenerate photon pairs is feasible, which can be spatially separable by using a dichroic mirror.
		
		\subsection{Generation of polarization-entangled photons with coexisting NBPM and QPM}
		With the photon pairs generated by the NBPM carefully characterized, the unpoled KTP crystal is replaced by a PPKTP crystal with a poling period of 2 mm to achieve NBPM and QPM simultaneously. The duty cycle of the periodic poling is designed to be $D\approx0.735$ (instead of the conventional $D=0.5$) for the first-order QPM so that the generation efficiencies of the photon pairs associated with the NBPM and QPM are equal. The crystal temperature was set at $25.8 ^{\circ}$C to generate non-degenerated biphotons with wavelengths of 1074.4 nm and 1080.8 nm while the wavelength of the pump is tuned to be 538.6 nm. The pump light is prepared from a frequency-doubled laser system.	An external cavity diode laser (ECDL) was frequency-tuned to be 1077.2 nm and its output power was amplified to be 700 mW.
		To generate the second harmonic (SH) light, the amplified light was collimated into a periodically poled lithium niobate (PPLN) crystal with a poling period of $7 \ \mu$m and length of 50 mm. With the temperature of the PPLN crystal set at about $160^{\circ}$C to achieve the type-I phase matching, SH light is generated at 538.6 nm with a maximum power of 7 mW. \\
  
By pumping the PPKTP crystal with the SH light at 4 mW, we generate polarization-entangled photons.
		The generated photon pairs pass through a compensation crystal (an unpoled KTP crystal with half length) in order to compensate the time delay between the signal and idler photons caused by the birefringence of the PPKTP crystal.
		A dichroic mirror (DM) with a cut-off slope of 5 nm is used to separate the signal photons (at 1074.4 nm) and idler photons (at 1080.8 nm) before they are sent into the polarization analyzers. Each analyzer is composed of a half-wave plate (HWP), a quarter-wave plate (QWP), and a PBS, which enable us to perform the test of the CHSH inequality and the QST as shown in Fig. 3(II). By attaching a TBF to a single mode fiber (SMF) before each SPCM to filter out the unwanted spatial or spectral modes, we measure the pair rate to be $R_c=439$ s$^{-1}$ with an accidental coincidence count rate of about $0.4$ s$^{-1}$. Among the measured pair rate, $222$ s$^{-1}$ results from the $\ket{H}_s\ket{V}_i$ photon pairs and $217$ s$^{-1}$ from the $\ket{V}_s\ket{H}_i$ pairs, indicating the ratio of the generation efficiencies of the NBPM and QPM to be 1 : 0.98. With the count rates of the signal and idler photons measured to be $R_s=2.18 \times 10^4$~s$^{-1}$ and $R_i=2.68 \times 10^4$~s$^{-1}$, respectively, it infers a generation rate of $R_sR_i/R_c \simeq 1.33 \times 10^6$ s$^{-1}$ or a brightness of $3.3 \times 10^5$ s$^{-1}$mW$^{-1}$. 
      \\\indent

      
        To verify the polarization entanglement, we measured the coincidence counts of the photon pairs while scanning the polarization angle $\theta_{i}$ of the idler photons with the polarization angle $\theta_{s}$ of the signal photons fixed in the diagonal (D) or anti-diagonal (A) polarization states. The polarization scanning is performed by rotating the HWP in the polarization analyzer of the idler photons with the angle $\theta_{\rm B} = \theta_i/2$. The experimental result is shown in Fig. 5. The visibilities of $98.7\%\pm0.14\%$ and $98.3\%\pm0.16\%$ for the signal photons in the D and A states, respectively, indicate that the signal and idler photons are highly entangled. To further confirm the entanglement, we verify the nonlocality of the photon pairs by testing the violation of the CHSH inequality \cite{Bell64,Clauser69},
        \begin{equation}
        S=\vert E(\theta _s,\theta _i)-E(\theta _s,\theta _i^{'})\vert+\vert E(\theta _s^{'},\theta _i)\vert+\vert E(\theta _s^{'},\theta _i^{'})\vert  \le 2,
        \end{equation} 
		where $E$ is the correlation function. The inequality holds for all states satisfying the local-hidden-variable (LHV) theorem \cite{Einstein35}. In contrast, a maximally entangled state gives $S=2\sqrt{2}$.
        The experimental \textit{S}-value of the photon pair generated in the present work is measured to be $S=2.763\pm0.009$, which violates 84 standard deviations beyond the classical limit predicted by the LHV theorem.\\
              \begin{figure}[h]
        	\graphicspath{{fig/}}
        	\includegraphics[width=0.6\textwidth]{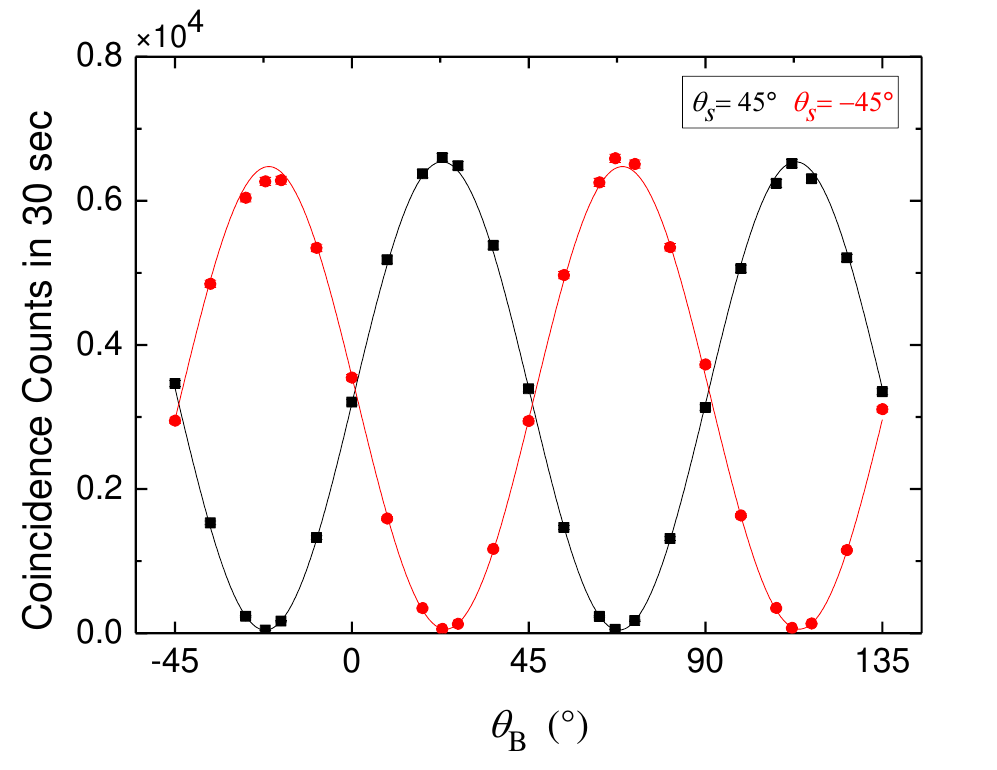}
        	\setlength{\belowcaptionskip}{-0cm}
        	\footnotesize
        	\centering
			\caption{Polarization correlations between the signal and idler photons. Coincidence counts of the photon pair are recorded by measuring the signal photons in the D state ($\theta_s=45^{\rm o}$, squares) or A state ($\theta_s=-45^{\rm o}$, circles) while scanning the polarization angle $\theta_{i}$ of the idler photons. The accidental coincidence counts are subtracted. The error bars representing the statistical errors are smaller than the size of the symbols. The visibilities with the D and A states are $98.7\%\pm0.14\%$ and $98.3\%\pm0.16\%$, respectively. 
        	}
        	\label{figure:induction loop}
        \end{figure}

        \begin{figure}[t]
        	\graphicspath{{fig/}}
        	\includegraphics[width=0.8\textwidth]{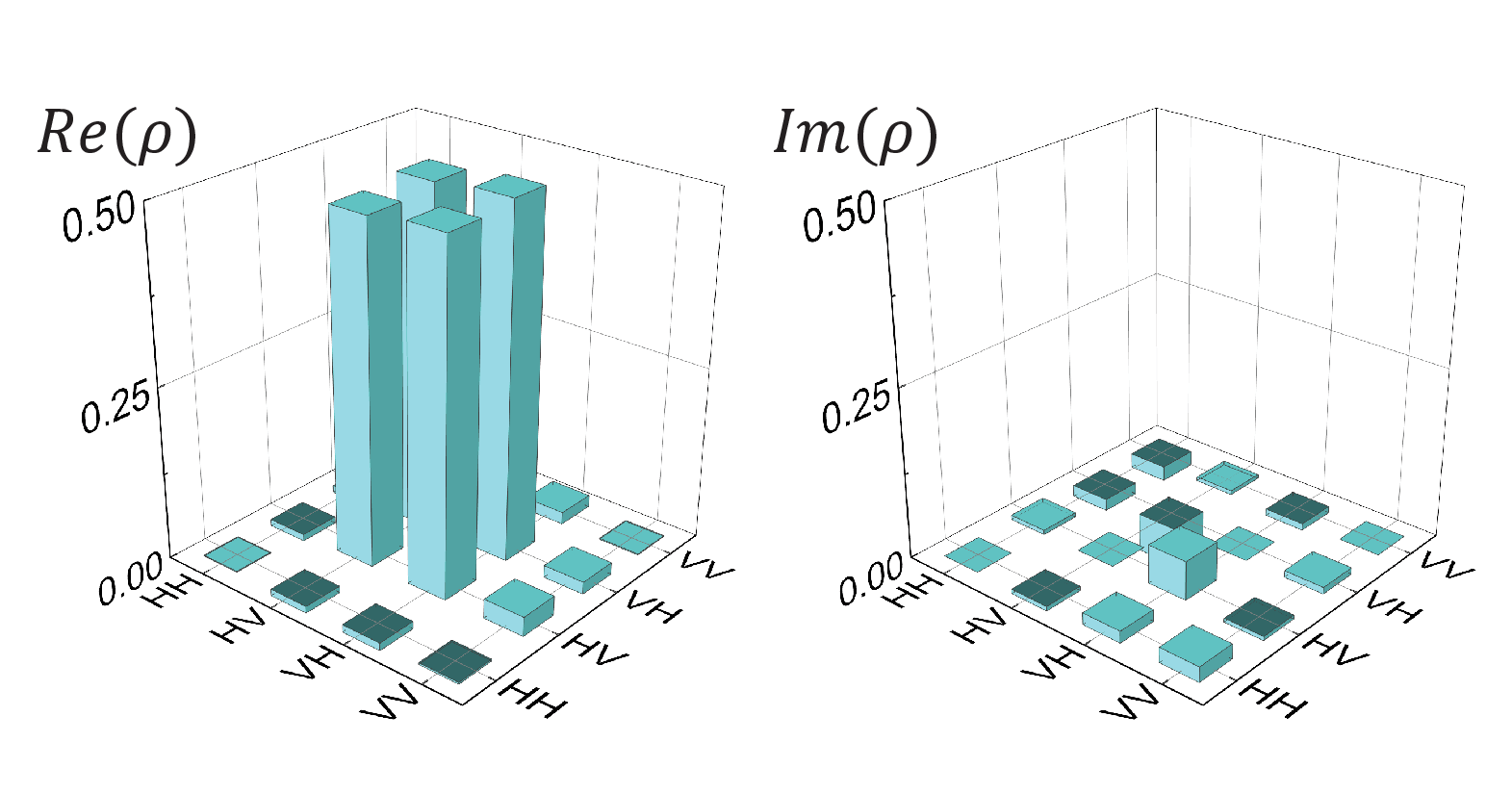}
        	\setlength{\belowcaptionskip}{-0cm}
        	\footnotesize
        	\centering
        	\caption{Quantum state tomography of the polarization-entangled photons generated by coexisting NBPM and QPM in a PPKTP crystal. The left panel shows the real part of the density matrix while the right panel shows the imaginary part. The accidental coincidence counts are subtracted when the density matrix is reconstructed. The concurrence is $C = 0.935$ and the fidelity comparing with a Bell state is $F = 0.998$.  
        	}
        	\label{figure:induction loop}
        \end{figure}   
        We also perform the QST to obtain the density matrix of the entangled state to show how similar the generated state is to a Bell state. Figure 6 shows the density matrix reconstructed from the measurements of QST, giving a fidelity $F=0.998$ comparing to the Bell state,
        \begin{equation}\ket{\psi}=\frac{1}{\sqrt{2}}(\ket{H}_s\ket{V}_i+\ket{V}_s\ket{H}_i),\end{equation}
        and a concurrence \cite{Wootters98} $C=0.935$. The high fidelity and concurrence clearly demonstrates the strong similarity between the generated state and the Bell state. The slight discrepancy possibly results from the imperfect alignment of the polarization analyzers or the non-ideal reflectance (transmission) of the signal (idler) photons at the dichroic mirror.
      \\\indent
      
		\section{Conclusion}
		In conclusion, we demonstrate a novel polarization-entangled photon source based on the coexisting NBPM and QPM in a PPKTP crystal with a large poling period (2 mm) and simple domain structure. The source does not require interferometer, delicate domain structures, post selection, or multiple crystals and is thus compact and insensitive to fabrication error. The entangled photons violate the CHSH inequality by 84 standard deviations, manifesting their strong nonlocality. The density matrix reconstructed from the QST shows that the state is close to a Bell state with a fidelity $F=0.998$ and concurrence $C=0.935$. Moreover, we show that the large poling period effectively alleviates the conversion efficiency drop resulting from the fabrication defect. Although PPKTP crystal is used in our experiment, one can also use the PPLN crystals which have high nonlinear coefficients (but the quality of the poling structures may be more difficult to control). Due to the collinear geometry of our source, a monolithic cavity \cite{Chuu12} can be fabricated to implement the resonant SPDC with long coherence time and narrow linewidth for efficient quantum communication \cite{Liu13,Kao23} or light-matter interaction at the single photon level \cite{Wu17,Cheng20}. The compact polarization-entangled photon source thus has potential applications for quantum technologies.

\section*{Funding}
National Science and Technology Council (110-2112-M-007-021-MY3). 

\section*{Disclosures}
The authors declare no conflicts of interest.

\section*{Data availability}
Data underlying the results presented in this paper are not publicly available at this time but may be obtained from the authors upon reasonable request.

\end{document}